\documentclass[review]{elsarticle}

\usepackage{lineno,hyperref}
\modulolinenumbers[100]
\journal{arXiv}

\usepackage{bm}
\usepackage{amsmath}
\usepackage{threeparttable}
\usepackage{booktabs}
\usepackage{multirow}
\usepackage{graphicx,amsmath, amssymb,lineno}
\usepackage{subfigure, graphicx}
\usepackage{amsmath,amssymb,amsthm}
\usepackage{algorithmic}
\usepackage{algorithm}
\usepackage{tabularx}
\usepackage{multirow}
\usepackage{makecell}
\usepackage{setspace}
\usepackage{appendix}

\usepackage{hyperref}

\biboptions{authoryear}

\begin{document}

\begin{frontmatter}
\title{Frequency-specific segregation and integration of human cerebral cortex: an intrinsic functional atlas}


\author[mymainaddress]{Zhiguo Luo}
\author[mymainaddress]{Ling-Li Zeng}
\author[mymainaddress]{Hui Shen}


\author[mymainaddress]{Dewen Hu\corref{mycorrespondingauthor}}
\cortext[mycorrespondingauthor]{Corresponding author}
\ead{dwhu@nudt.edu.cn}

\address[mymainaddress]{College of Intelligence Science and Technology, National University of Defense Technology, Changsha, Hunan 410073, China}

\begin{abstract}
The frequency-specific coupling mechanism of the functional human brain networks underpins its complex cognitive and behavioral functions. Nevertheless, it is not well unveiled what are the frequency-specific subdivisions and network topologies of the human brain. In this study, we estimated functional connectivity of the human cerebral cortex using spectral connection, and conducted frequency-specific parcellation using eigen-clustering and gradient-based methods, and then explored their topological structures. 7T fMRI data of 184 subjects in the HCP dataset were used for parcellation and exploring the topological properties of the functional networks, and 3T fMRI data of another 890 subjects were used to confirm the stability of the frequency-specific topologies. Seven to ten functional networks were stably integrated by two to four dissociable hub categories at specific frequencies, and we proposed an intrinsic functional atlas containing 456 parcels according to the parcellations across frequencies. The results revealed that the functional networks contained stable frequency-specific topologies, which may imply more abundant roles of the functional units and more complex interactions among them.
Parcellation results generated from 7T fMRI data of 184 participants are publicly available \href{https://github.com/luozhuxi/Frequency-parcellation}{Github\_Frequency-parcellation}.

\end{abstract}

\begin{keyword}
 multi-taper coherence  \sep connector hubs   \sep eigen-clustering \sep topology  \sep parcellation
\end{keyword}

\end{frontmatter}

\linenumbers

\section{Introduction}
The tremendous progress of neuroimaging technology has greatly promoted the research of the human brain in recent years \citep{Bullmore-2009-p186-198, Buechel-1998-p947-957, Deco-2011-p43-56, Baron-Cohen-1999-p1891-1898, Biswal-1995-p537-541, Dosenbach-2010-p1358-1361}. Resting-state functional magnetic resonance imaging (rs-fMRI) can measure the spontaneous low-frequency oscillations of blood oxygen level-dependent (BOLD) signals, and resting-state functional connectivity can comprehensively and non-invasively measure the interaction of neural activity between brain areas without distance constraints  \citep{Bullmore-2009-p186-198}. This has aroused great interest in the neuroscience and medical community because more and more evidence shows that the pathological conditions of many cortex diseases are related to abnormal functional connectivities between specific cortex regions \citep{Church-2008-p225-238, Liu-2008-p1648-1656, Luo-2014-p431-441, Seeley-2009-p42-52, Wang-2006-p496-504, Baudrexel-2011-p1728-1738}.

Researchers have used various methods to study functional connectivity on fMRI data, and one of the most successful methods is graph theory. In the framework of graph theory, a complex system is regarded as a graph composed of nodes and edges, which corresponds to ROIs and functional connectivities, respectively. The application of graph theory in the research of functional connectivity has been relatively mature, such as successfully identifying functional networks and sub-networks, and successfully portraying the hierarchical organization of the human brain \citep{Power-2011-p665-678}. The functional segregation and integration in the human brain indicate that functional units or networks may either individually respond to a certain task stimulus, or interact with each other to complete intricate neural activities. At the same time, studies have shown that neural signals may contain effective frequency-specific components, and there is related fluctuations of neural signals in different brain areas in multiple time scales \citep{Eguiluz-2005-p18102-18102, Lei-2013-p52-57,Zuo-2010-p1432-1445, He-2011-p13786-13795,Baria-2011-p7910-7919}. However, there is still a lack of systematic research on the interaction mechanism among functional units in separate frequency bands. Therefore, we attempt to study the segregation and integration mechanism of the functional human brain networks by clarifying their topological structures in a frequency perspective.

The prerequisite for accurately constructing topology of the functional human brain is a proper functional parcellation. When the functional network is modeled as a graph, it will be distorted if the nodes cannot be completely and accurately represent the functional units \citep{Wig-2011-p126-146, Smith-2011-p875-891, Butts-2009-p414-416}. The use of existing atlas or templates based on static functional connectivity may distort the topological structure of frequency-specific functional network, therefore, frequency-specific functional parcellation is needed. 

The resting-state functional network was once integrated by three types of connector hubs: control-default, control-control, and control-processing \citep{Gordon-2018-p4-1695}. We therefore hypothesize that if the frequency-specific topological structures of the functional networks are various. The types of connector hubs may be different across frequencies.
We have two purposes in this study. First, we want to conduct frequency-specific functional parcellation of the human cerebral cortex, and analyze the similarities and differences between them and the static functional networks. Secondly, we want to build topology models of these frequency-specific functional networks and analyze their unique roles in neural activity.

\section{Materials and methods}
\subsection{ Data acquisition and preprocessing}
All fMRI data are from the HCP dataset, among which 3T fMRI data are described in detail in our previous study \citep{Luo-2019-p-269-282}. The scanning direction of the 7T fMRI data is AP (anterior-to-posterior)/PA (posterior-to-anterior) \citep{Vu-2017-p23-32}, which is different from the scanning direction of 3T fMRI data (LR/RL), and 7T fMRI data greatly improves the signal-to-noise ratio. Both 3T and 7T data are preprocessed by the standard HCP preprocessing pipelines and registered to the 32k CIFTI space. All 184 subjects in the HCP 7T dataset were selected for parcellation and network topology exploring in this study, and 890 subjects in the 3T dataset excluding those who passed the 7T scan at the same time were used for repeatability experiments on the network topology. We removed the first 20 frames of 3T fMRI data (TR=0.72s) and the first 15 frames of 7T fMRI data (TR=1s) to ensure magnetic saturation and performed detrending to calculate spectral connection (coherence) between vertices at discrete frequency bands. The fMRI data were further band-pass filtering (0.01$\sim$0.08 $Hz$) to calculate the temporal connection (correlation) between vertices. Since the sample size of 7T data is small, we selected all the data whose scanning direction are PA, while the sample size of 3T data is larger, we only selected REST1\_LR data.

\subsection{Frequency-specific parcellation of the human cerebral cortex}%
\begin{figure}[ht]
	\centering
	\includegraphics[trim=0 485 0 0,scale=1.0,width=12cm]{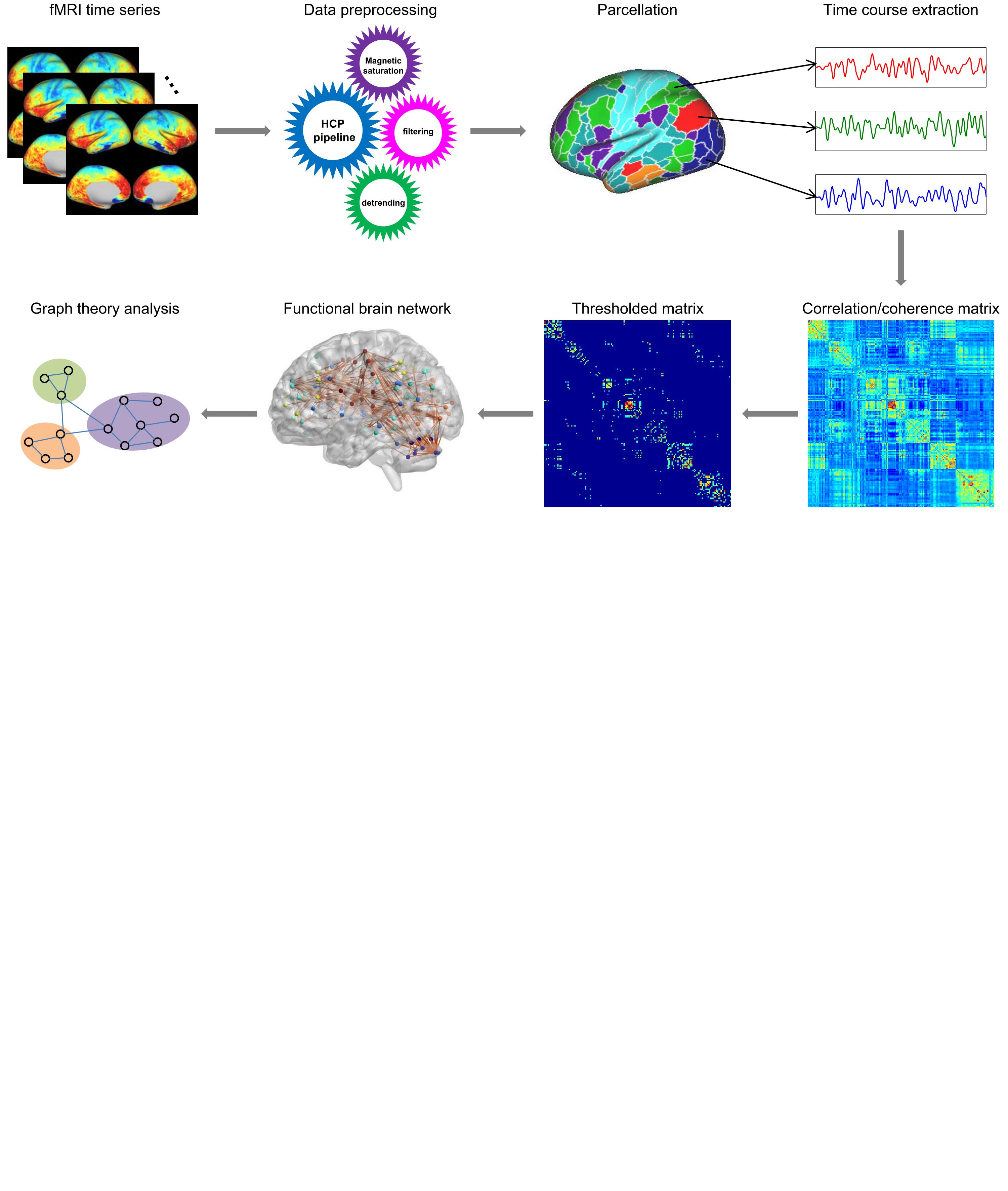}
	\caption{Framework of whole-cortical parcellation and graph theory analysis of functional functional network. The cerebral cortex is parcellated using fMRI data undergone a series of preprocessing, and then we obtained the average time-series from parcels and calculate the functional connectivities between them. To reduce the complexity of the data, the functional connectivity matrix is sparsely processed with a series of thresholds. And then we combine the results of the network and area parcellation to obtain the cortical functional network, and perform graph theory analysis on this basis.}
	\label{fig:topology_pipeline}
\end{figure}

The framework of this study is shown in Figure \ref{fig:topology_pipeline}. After the preprocessing of the HCP fMRI data, we first conducted the group-level static and frequency-specific parcellation of the human cerebral cortex. In detail, for each cortical vertex of each subject (64984 vertices in total, among which 59412 vertices contain valid time-series), we calculate the temporal connection (correlation) and spectral connection (coherence) between them and the 360 ROIs from \citep{Glasser-2016-p171-178} to obtain one correlation matrix and eight coherence matrices (0.01$\sim$0.08 $Hz$, step=0.01 $Hz$). The coherence of two time-series $x, y$ at frequency $\lambda$ is defined as:
\begin{equation}
Co{h_{xy}}(\lambda) = \frac{{{{\left| {{f_{xy}}(\lambda )} \right|}^2}}}{{\left[ {{f_ {xx}}(\lambda ){f_{yy}}(\lambda )} \right]}}
\end{equation}
where ${f_{xy}}(\lambda )$ is the cross spectrum of $x$ and $y$, ${f_{xx}}(\lambda )$ is the power spectrum of $x$, and ${f_ {yy}}(\lambda )$ is the power spectrum of $y$ \citep{Brillinger-1981-p-,Mueller-2001-p347-356,Curtis-2005-p177-183}.  We detected the spectrum component using the multi-taper spectrum estimation algorithm. After that, those functional connectivity matrices were z-transformed and group-averaged across subjects. In this way, we have obtained nine group-averaged functional connectivity matrices of 184 people with the dimension of 59412 $\times$ 360 (one correlation matrix and eight coherence matrices), and the whole cortical parcellation was conducted based on them.

Considering the hierarchical organization of the functional networks \citep{Zhou-2006-p-238103-238103}, it is not easy to depict the functional networks and parcels simultaneously based on functional connectivity without any prior knowledge or constraint.  In this study, we got functional networks using the global similarity approach (clustering) and parcels using the local gradient approach.

\subsubsection{network identification via clustering}
To get frequency-specific functional networks, it is inappropriate to determine the number of networks subjectively. We used our previously proposed eigen-clustering method as it can automatically determine the cluster number \citep{Luo-2019-p-269-282}, and made some modifications for network parcellation.

 Since the eigen-clustering algorithm needs to select the optimal clustering result from several candidates for brain area  parcellation, we set three selection standards: number of scatters, cluster validity index, and symmetry index  \citep{Luo-2019-p-269-282}, which are not entirely applicable for the whole functional network parcellation. For example, to get spatially connected parcels, it is reasonable to use the number of scatters as a metric for the quality of the parcellation results. Nevertheless, the functional network allows discrete regions to have the same label for functional integration.
For another example,  we found symmetry is the characteristic of the precuneus in the practice of parcellation, which may not apply to other brain regions, and accumulating evidences pointed out that the distribution of functional networks presents the characteristics of lateralization \citep{Schaefer-2018-p3095-3114, Craddock-2011-p1914-1928, Shen-2013-p403-415, Baldassano-2015-p784-784, Gordon-2016-p288-303}, so symmetry is also not quite reasonable to be used as an index to measure the results of network parcellation. The only indicator that can continue to be applied is the cluster validity index, but we believe that it can be used as one of the criteria to help select the optimal clustering result, and when it is used as the only selection criterion, sometimes a poorer result may be selected. An example is in Table 1 of \cite{Luo-2019-p-269-282}.  From the practice of the precuneus parcellation using  eigen-clustering, we found that if all the Laplacian eigenvectors with dimension of $d$ = 3$\sim$15 are concatenated to form a new feature vector (vector dimension $d = \sum\limits_{k = 3}^{15} k = 117$), and the density-peaks clustering \citep{Rodriguez-2014-p1492-1496} was conducted based on the distance matrix of the new feature vectors. The experimental results showed that the parcellation results obtained by the improved eigen-clustering are similar to the optimal result obtained by the original version. Therefore, in this study, the network parcellation was conducted using a developed eigen-clustering algorithm.

\subsubsection{parcel creation using boundary maps}
Parcel creation was conducted using a gradient-based cortical subdivision algorithm proposed by \cite{Cohen-2008-p45-57}. We first calculate the RSFC map similarity matrix by calculating the pairwise spatial correlations among all vertices of both hemispheres. Because the Laplacian eigenmaps helps to reduce high-dimensional noise, and for the consistency of the global (network) and local (parcel) parcellation, the functional similarity between vertices was measured as the Pearson correlation coefficient of the concatenated Laplacian eigenvectors ($d=117$).
For each column of the similarity matrix ($64k \times 64k$), we use Connectome Workbench to calculate the first spatial derivative, resulting in $64k$ gradient maps for each hemisphere, then the watershed algorithm was used to get $64k$ boundary maps and those boundary maps were averaged and the watershed algorithm was used again to identify the final boundary map of the averaged boundary map of each hemisphere.
 
 Parcels were created from the final obtained boundary map. First, local minima were defined as vertices with values smaller than their neighbors within 3 vertices, then parcels were grown from those local minima using the watershed algorithm until they met each other. The resulting parcels were then merged based on two standards referring to \citep{Gordon-2016-p288-303, Gordon-2017-p7-807, Gordon-2018-p4-1695}. First, if the parcel contains fewer than 30 vertices, it will be merged to its neighbors with the minimum median boundary value between them. Second, if the minimum median boundary value of two neighboring parcels were less than the $60^{th}$ percentile of the whole boundary map's values, they will be merged. For the utmost correspondence of the global and local parcellation, merging was not conducted across networks. In this way, each vertex of the cerebral cortex has a network label and a parcel label.

\subsection{hub identification and clustering}
Hub identification and clustering were conducted referring to \citep{Gordon-2018-p4-1695}. For each participant, the hub is defined according to the participation coefficient (PC) metric. First, the parcel signal was defined as the average time-series of all vertices in this parcel, and functional connectivities (spectral or temporal connection) among those parcels were calculated to create eight coherence matrices and one correlation matrix. If the geodetic distance between the centroids of two parcels is less than 15 vertices (about 30 $mm$), the corresponding element in the correlation/coherence matrix is set to 0. Referring to \cite{Gordon-2018-p4-1695}, we set a series of graph density thresholds (0.3\%$\sim$0.5\%, in 0.1\% increments; and 0.5\%$\sim$5\%, in 0.5\% increments) to retain the largest elements in the matrix. For each thresholded matrix, the PC value of parcel $i$ was calculated as follows:
\begin{equation}
{P_i} = 1-{\sum\limits_{m \in M} {(\frac{{{K_i}(m)}}{{{K_i}}})} ^2}
\label{equ:PC}
\end{equation}
where $M$ is the set of all networks, $K_i$ is the total number of edges connected to parcel $i$, that is, the degree of this parcel. ${{K_i}(m)}$ is the number of edges connected to parcel $i$ within network $m$ \citep{Lynch-2018-p3912-3921}. Generally speaking, a parcel with a lower degree will not be a candidate hub, and on the other hand, a small degree $K_i$ may result in a large $P_i$ as it is used as the denominator, so we set the PC value of a parcel to be 0 if its degree is less than the first quartile of the degrees of all parcels. After that, PC values were converted to percentiles for normalization and averaged across density thresholds. Parcels with the top $80^{th}$ average percentile PC values were defined as connector hubs \citep{Bertolero-2015-p6798-6807, Gratton-2016-p1276-1288, Gordon-2018-p4-1695}.

As for hub clustering, we first calculated the functional connectivities between hubs and all functional networks at the individual level, and then hubs were clustered based on their network connectivity profiles across all individuals. Unlike traditional hub clustering methods that used the Louvain algorithm,  which needed about 1000 times of repetitions to get consensus results \citep{Gordon-2018-p4-1695}, we used the density-peaks clustering method \citep{Rodriguez-2014-p1492-1496}, which could determine the cluster number automatically and get stable results.

\section{Results}
\subsection{Parcellation and hub clustering of the human cerebral cortex at specific frequency bands}
\begin{figure}[ht]
	\centering
	\includegraphics[trim=0 180 0 0,scale=1.0,width=12cm]{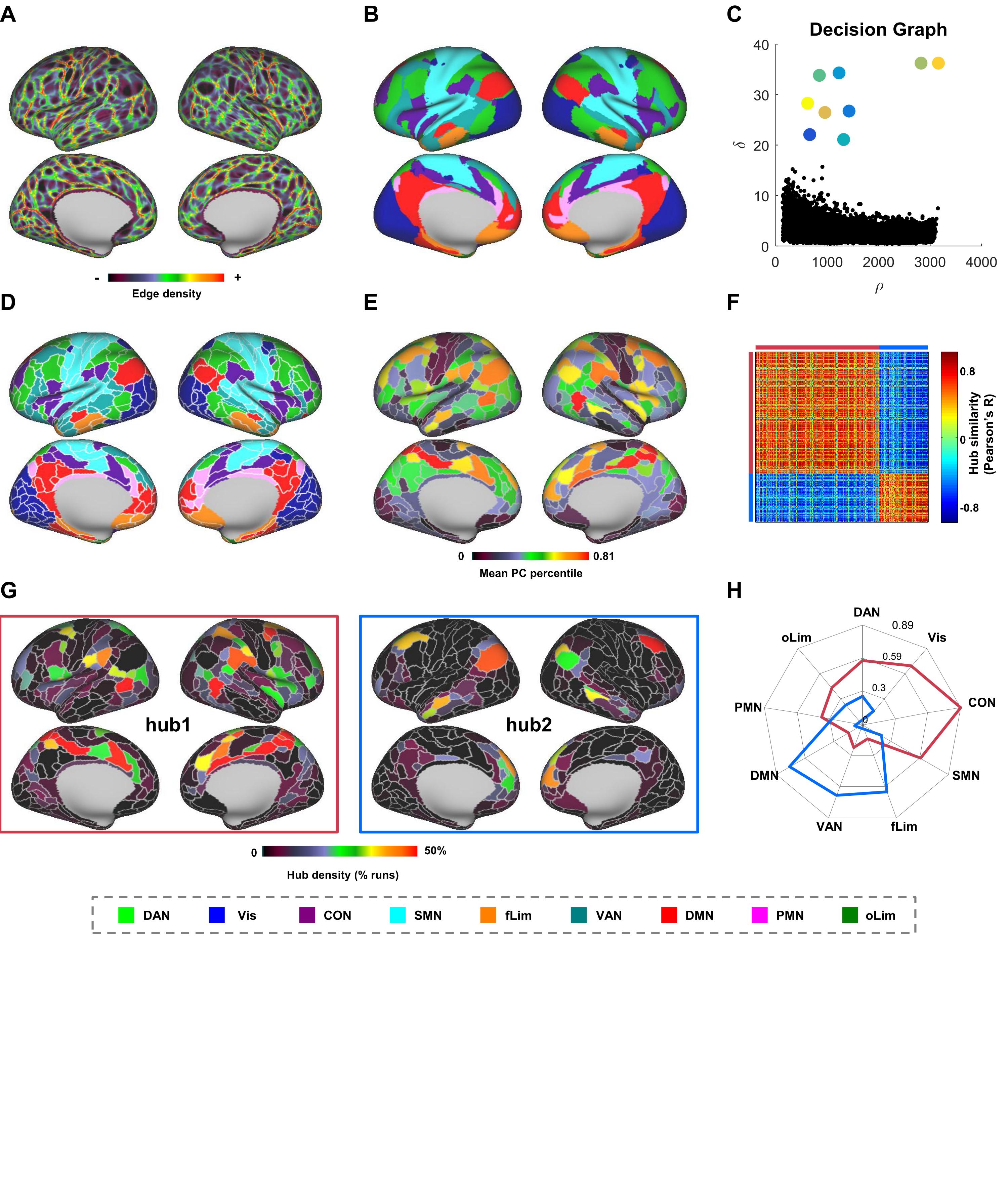}
	\caption{Static cortical parcellation and hub clustering. (A) Group average edge density map based on static correlation; (B) Eigen-clustering network parcellation results; (C) Eigen-clustering decision graph; (D) Whole cortical parcellation results; (E) Average PC percentile; (F) Hub clustering result, the border color indicates the hub set; (G) The spatial distribution of the hub set and the border color represents the hub set; (H) Average network connectivity of each set of hubs. The line color indicates the hub set.}
	\label{fig:static}
\end{figure}

The results of functional parcellation and hub clustering based on correlation are shown in Figure \ref{fig:static}. The whole cerebral cortex is clustered into 9 networks: the dorsal attention network (DAN), the visual network (Vis), the cingulo-opercular network (CON), the sensory-motor network (SMN), the fronto-limbic network (fLim), the ventral attention network (VAN), the default-mode network (DMN), the parietal memory network (PMN) and the occipito-limbic network (oLim), the network distribution presents a strong symmetry (the symmetry index $SI$=82.42\%). 
We can see a Vis feature in the lateral parietal cortex, just like \citep{Gordon-2017-p7-807}. 
Combined with the edge density map, the cerebral cortex is divided into 417 parcels, including 206 parcels in the left hemisphere and 211 parcels in the right hemisphere. We calculated the PC values of the parcels at the individual level and converted them to percentiles and averaged across individuals. We found that the parcels with the largest PC values were mainly located in the anterior precuneus, the dorsal posterior cingulate, the middle posterior lobe, and the upper-middle frontal gyrus. The clustering results showed that there were two types of connector hubs in the static network. The first type accounted for 71.65\% of all the hubs, and they mainly located in the anterior precuneus, the dorsal posterior cingulate, the anterior cingulate, the supramarginal gyrus, and the posterior meditemporal gyrus, showing strong correlation with the visual network, the cingulo-opercular network, and the sensory-motor network. The second type of hubs accounted for 28.35\% and mainly located in the anterior prefrontal lobe, the anterior meditemporal gyrus, the middle posterior lobe, and the upper-middle frontal gyrus, showing a strong correlation with the fronto-limbic network, the ventral attention network, and the default-mode network. The two types of hubs showed opposite connectivity with the fronto-limbic network and the occipito-limbic network, which proves the rationality of the network parcellation from one side.

\begin{figure}[ht]
	\centering
	\includegraphics[trim=0 180 0 0,scale=1.0,width=12cm]{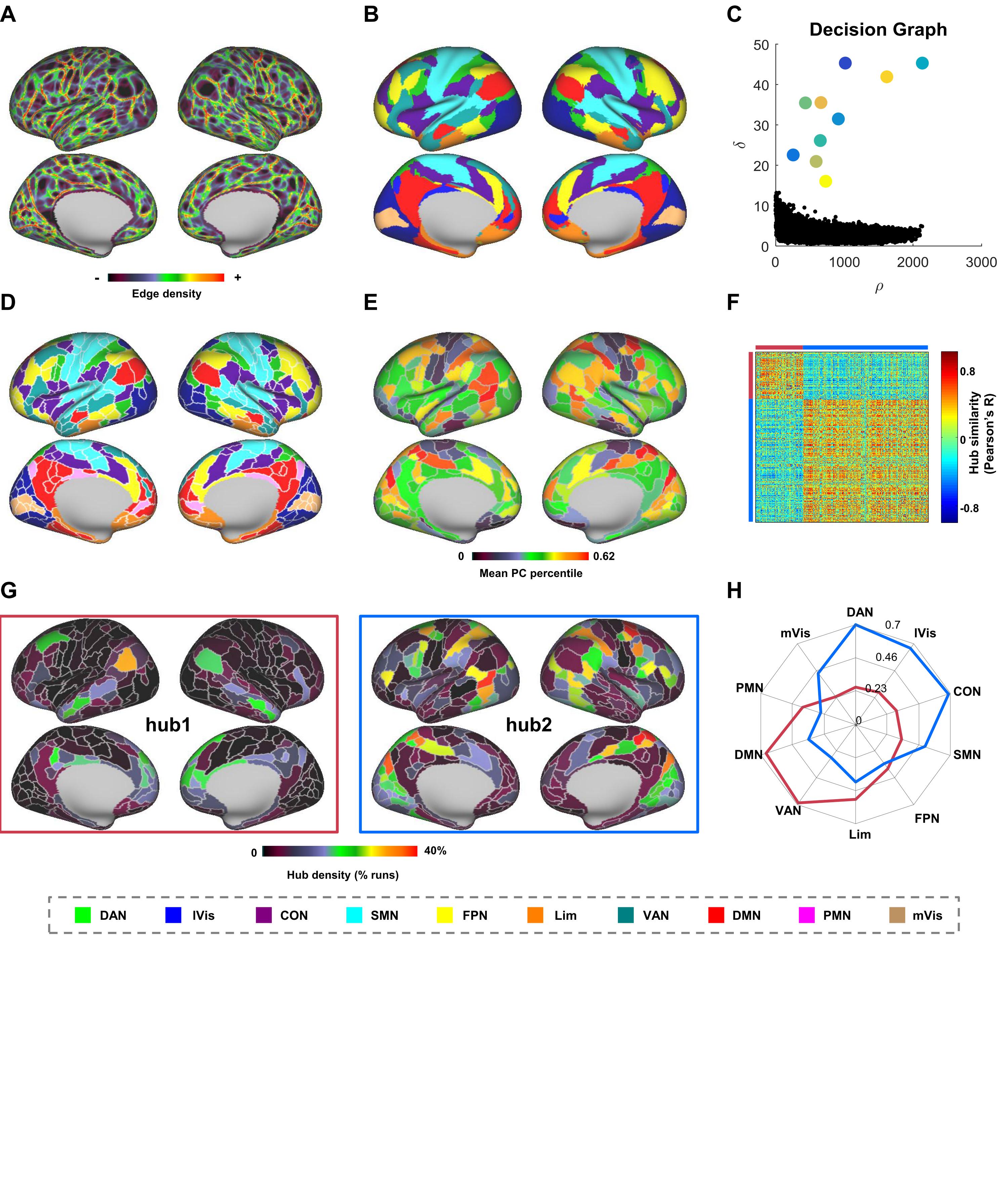}
	\caption{Cortical parcellation and hub clustering at 0.01 $Hz$. (A) Group average edge density map based on coherence at 0.01 $Hz$; (B) Eigen-clustering network parcellation results; (C) Eigen-clustering decision graph; (D) Whole cortical parcellation results; (E) Average PC percentile; (F) Hub clustering result, the border color indicates the hub set; (G) The spatial distribution of the hub set and the border color represents the hub set; (H) Average network connectivity of each set of hubs. The line color indicates the hub set.}
	\label{fig:fre1}
\end{figure}

The results of functional parcellation and hub clustering based on coherence at 0.01 $Hz$ are shown in Figure \ref{fig:fre1}. The entire cerebral cortex is clustered into 10 networks. Compared with the correlation-based static network parcellation, the visual network is subdivided into the lateral visual network (lVis) and medial visual network (mVis), and the fronto-parietal control network (FPN) is separated from the dorsal attention network, and the fronto-limbic network and the occipito-limbic network are merged into a limbic network (Lim). Coherence network distribution at 0.01 $Hz$ also shows strong symmetry ($SI$=82.47\%). Combined with the edge density map, the cerebral cortex is subdivided into 391 parcels, including 201 parcels on the left and 190 parcels on the right. We found that the parcels with the largest PC values were mainly located in the anterior and posterior precuneus, the dorsal posterior cingulate, and the anterior and posterior middle temporal gyrus. The clustering results show that there are two types of connector hubs in 0.01 $Hz$ coherence functional network. The first type of hubs accounted for 27.58\% and mainly located in the anterior frontal lobe, the ventral posterior cingulate, the middle of the precuneus, the anterior parietal lobe, and the posterior middle temporal gyrus. They showed intense coherence with Lim, VAN, and DMN. The second type of hubs accounted for 72.42\% and mainly located in the anterior and posterior precuneus and the dorsal posterior cingulate, and showed strong coherence with DAN, lVis, CON, and SMN. The proportion and distribution of hubs at 0.01 $Hz$ and its network connection mode is roughly the same as hubs of the static network.

\begin{figure}[ht]
	\centering
	\includegraphics[trim=0 50 0 0,scale=1.0,width=12cm]{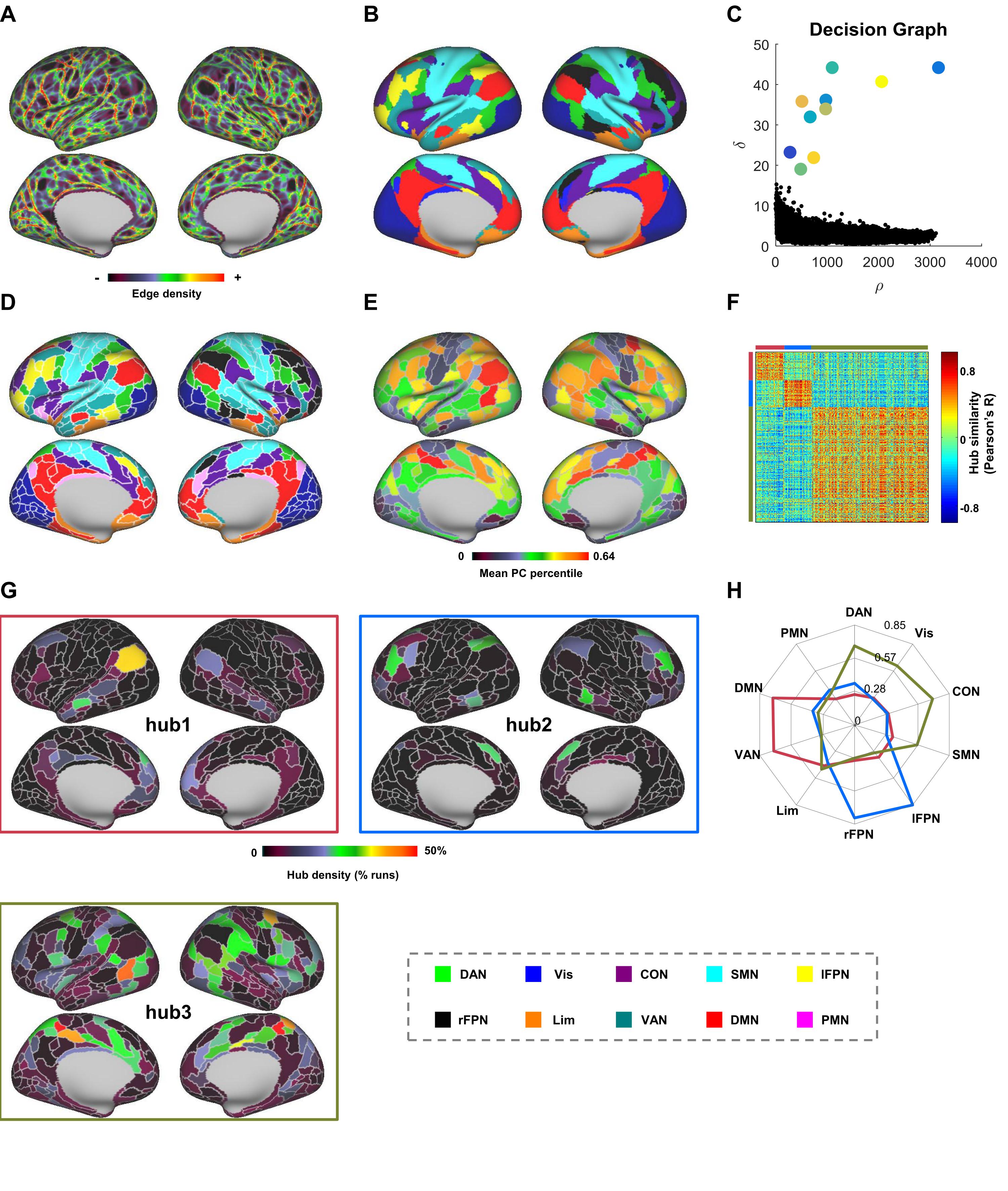}
	\caption{Cortical parcellation and hub clustering at 0.02 $Hz$. (A) Group average edge density map based on coherence at 0.02 $Hz$; (B) Eigen-clustering network parcellation results; (C) Eigen-clustering decision graph; (D) Whole cortical parcellation results; (E) Average PC percentile; (F) Hub clustering result, the border color indicates the hub set; (G) The spatial distribution of the hub set and the border color represents the hub set; (H) Average network connectivity of each set of hubs. The line color indicates the hub set.}
	\label{fig:fre2}
\end{figure}

The results of functional parcellation and hub clustering based on coherence at 0.02 $Hz$ are shown in Figure \ref{fig:fre2}. The entire cerebral cortex is clustered into 10 networks. Compared with networks at 0.01 $Hz$, the visual network is not further subdivided, while the fronto-parietal control network is split into left and right parts, showing obvious lateralization. Affected by this, the symmetry of 0.02 $Hz$ coherence network was reduced ($SI$=76.60\%). Combined with the edge density map, the cerebral cortex is subdivided into 407 parcels, including 204 parcels on the left and 203 parcels on the right. We found that parcels with the largest PC values were mainly located in the left posterior lobe, the left posterior middle temporal gyrus, the anterior precuneus, the dorsal posterior cingulate, and the right precentral gyrus. The clustering results show that there are three types of connector hubs in 0.02 $Hz$ coherence network. The first type of hubs accounted for 16.72\% and mainly located in the posterior lobe, the anterior prefrontal lobe, and the anterior middle temporal gyrus, showing strong coherence with DMN and VAN. The second type of hubs accounted for 15.60\% and mainly located in the middle temporal gyrus, the middle prefrontal lobe, the upper and lower part of the middle frontal gyrus, and showed strong coherence with bilateral FPNs. Although left and right FPN showed obvious lateralization in the network parcellation, the second type of hubs showed significantly higher coherence with both of them than other networks. Six among eight coherence networks contained lateralized left and right FPNs and a type of connector hubs showing high coherence with both of them. The third type of hubs accounts for 67.68\% and mainly located in the anterior precuneus, the posterior cingulate, the posterior middle temporal gyrus, the anterior precentral gyrus, and the anterior cingulate, showing strong coherence with DAN, Vis, CON, and SMN.

\begin{figure}[ht]
	\centering
	\includegraphics[trim=0 50 0 0,scale=1.0,width=12cm]{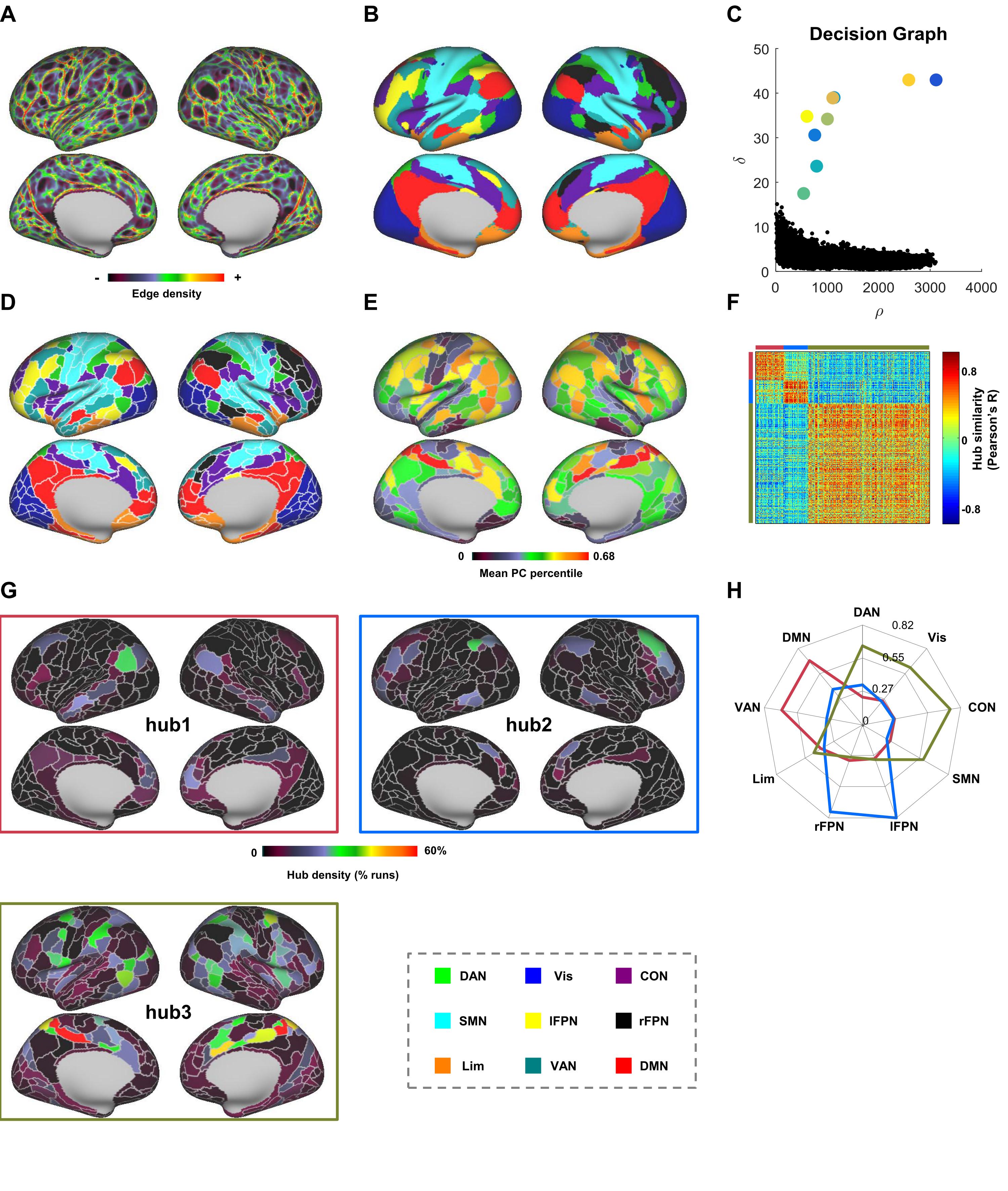}
	\caption{Cortical parcellation and hub clustering at 0.03 $Hz$. (A) Group average edge density map based on coherence at 0.03 $Hz$; (B) Eigen-clustering network parcellation results; (C) Eigen-clustering decision graph; (D) Whole cortical parcellation results; (E) Average PC percentile; (F) Hub clustering result, the border color indicates the hub set; (G) The spatial distribution of the hub set and the border color represents the hub set; (H) Average network connectivity of each set of hubs. The line color indicates the hub set.}
	\label{fig:fre3}
\end{figure}

\begin{figure}[ht]
	\centering
	\includegraphics[trim=0 50 0 0,scale=1.0,width=12cm]{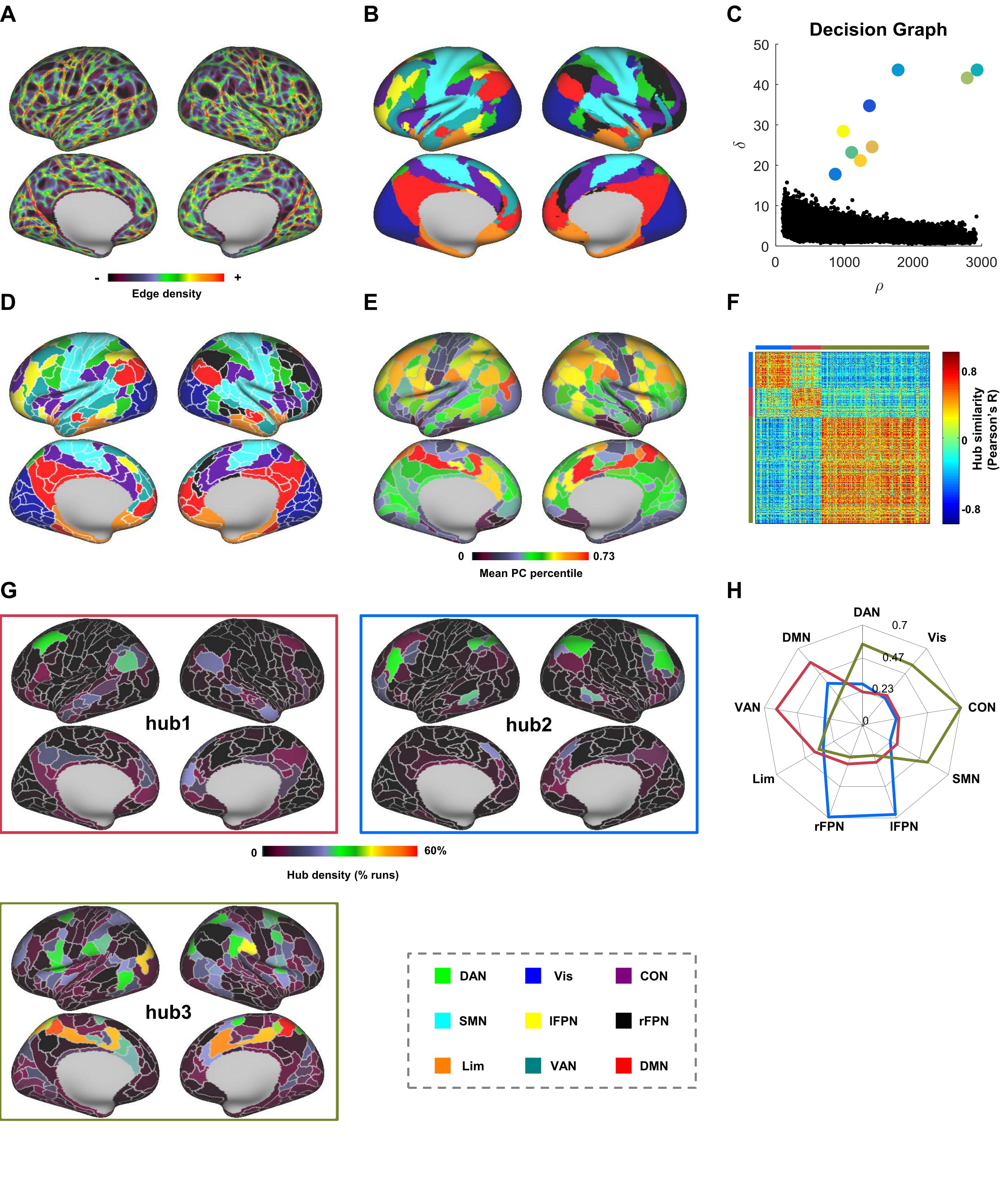}
	\caption{Cortical parcellation and hub clustering at 0.06 $Hz$. (A) Group average edge density map based on coherence at 0.06 $Hz$; (B) Eigen-clustering network parcellation results; (C) Eigen-clustering decision graph; (D) Whole cortical parcellation results; (E) Average PC percentile; (F) Hub clustering result, the border color indicates the hub set; (G) The spatial distribution of the hub set and the border color represents the hub set; (H) Average network connectivity of each set of hubs. The line color indicates the hub set.}
	\label{fig:fre6}
\end{figure}

The results of functional parcellation and hub clustering based on coherence at the frequency band of 0.03$\sim$0.06 $Hz$ are roughly the same, as shown in Figure \ref{fig:fre3}$\sim$\ref{fig:fre6}. The entire cerebral cortex is clustered into 9 networks: DAN, Vis, CON, SMN, lFPN, rFPN, Lim, VAN, and DMN. Detailed parcels of the cerebral cortex are as follows: 205 parcels on the left and 204 parcels on the right at 0.03 $Hz$; 205 parcels on the left and 216 parcels on the right at 0.04 $Hz$; 209 parcels on the left and 208 parcels on the right at 0.05 $Hz$; 204 parcels on the left and 211 parcels on the right at 0.06 $Hz$. The parcels with the largest PC value were mainly located in the anterior precuneus, the anterior and posterior cingulate, the posterior middle temporal gyrus, the insula, the superior parietal lobule, and the supramaginal gyrus. There are 3 types of hubs in the functional network. The first type of hubs is mainly situated in the anterior prefrontal lobe, the posterior lobe, and the middle temporal gyrus. They have strong coherence with DMN and VAN. This type of hub occupied 16.31\%, 18.28\%, 15.71\%, and 17.00\% of all the hubs at 0.03$\sim$0.06 $Hz$, respectively. The second type of hubs are mainly located in the upper posterior lobe, the middle frontal gyrus, the prefrontal lobe, and the meditemporal gyrus, showing strong coherence with bilateral FPNs. This type of hubs occupied 13.76\%, 19.68\%, 23.99\%, and 20.92\% of all the hubs at 0.03$\sim$0.06 $Hz$, respectively. The third type of hubs are mainly located in the anterior precuneus, the upper posterior cingulate, the anterior cingulate, the posterior frontal lobe, the insula, the superior temporal sulcus, and the middle temporal gyrus, showing strong coherence with DAN, Vis, CON, and SMN. This type of hub occupied 69.93\%, 62.05\%, 60.30\%, and 62.07\% of all the hubs at 0.03$\sim$0.06 $Hz$, respectively.

\begin{figure}[ht]
	\centering
	\includegraphics[trim=0 50 0 0,scale=1.0,width=12cm]{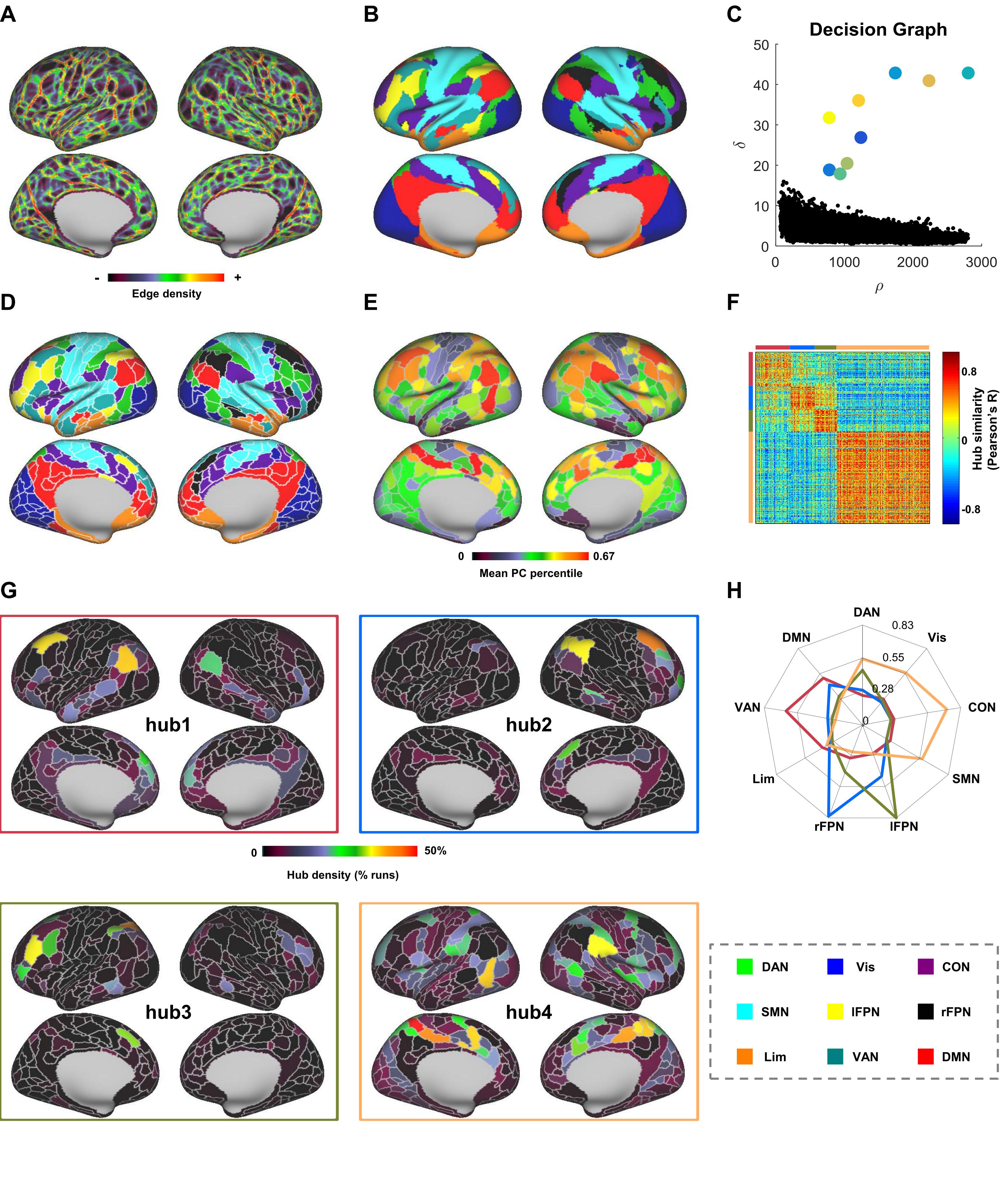}
	\caption{Cortical parcellation and hub clustering at 0.07 $Hz$. (A) Group average edge density map based on coherence at 0.07 $Hz$; (B) Eigen-clustering network parcellation results; (C) Eigen-clustering decision graph; (D) Whole cortical parcellation results; (E) Average PC percentile; (F) Hub clustering result, the border color indicates the hub set; (G) The spatial distribution of the hub set and the border color represents the hub set; (H) Average network connectivity of each set of hubs. The line color indicates the hub set.}
	\label{fig:fre7}
\end{figure}

\begin{figure}[ht]
	\centering
	\includegraphics[trim=0 180 0 0,scale=1.0,width=12cm]{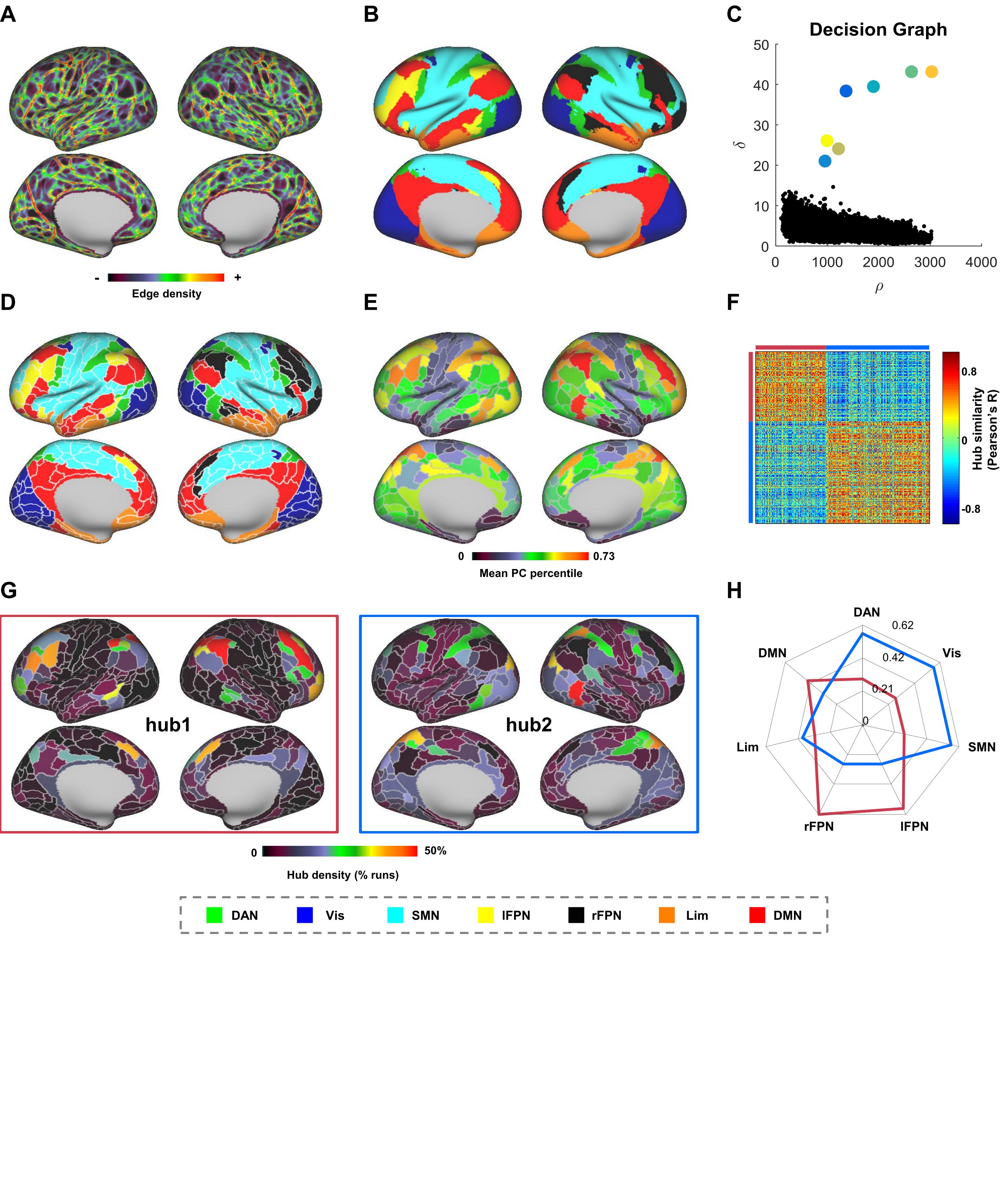}
	\caption{Cortical parcellation and hub clustering at 0.08 $Hz$. (A) Group average edge density map based on coherence at 0.08 $Hz$; (B) Eigen-clustering network parcellation results; (C) Eigen-clustering decision graph; (D) Whole cortical parcellation results; (E) Average PC percentile; (F) Hub clustering result, the border color indicates the hub set; (G) The spatial distribution of the hub set and the border color represents the hub set; (H) Average network connectivity of each set of hubs. The line color indicates the hub set.}
	\label{fig:fre8}
\end{figure}

The results of functional parcellation and hub clustering based on coherence at 0.07 $Hz$ are shown in Figure \ref{fig:fre7}. The entire cerebral cortex was also clustered into 9 networks. Combined with the edge density map, the cerebral cortex is divided into 427 parcels, including 210 parcels on the left and 217 parcels on the right. Parcels with the largest PC value are located in the junction of the precuneus, the posterior cingulate, and the paracentral gyrus, the middle frontal gyrus, the left posterior lobe, the left meditemporal gyrus and the supramaginal gyrus. The clustering results show that there are four types of connector hubs in the functional network at 0.07 $Hz$. The results of functional parcellation and hub clustering based on the coherence at 0.08 $Hz$ are shown in Figure \ref{fig:fre8}. The whole cerebral cortex is clustered into 7 networks. Compared with the 9 functional networks at the frequency band of 0.03$\sim$0.07 $Hz$, VAN and DMN merged into a new network, and CON and SMN merged into a new network. Combined with the edge density map, the cerebral cortex is subdivided into 429 parcels, including 218 parcels on the left and 211 parcels on the right. Parcels with the largest PC value are mainly located in the right hemisphere, including the right middle temporal gyrus and the posterior inferior temporal gyrus, the upper and lower right middle frontal gyrus, the right superior marginal gyrus, and the front parietal junction. The clustering results show that there are two types of connector hubs in 0.08 $Hz$ coherence functional network. The first type of hubs accounted for 40.53\% and mainly located in the middle of the bilateral prefrontal lobe, the bilateral parietal frontal lobe, the middle frontal gyrus. They showed strong coherence with the bilateral FPN and DMN. The second type of hubs accounted for 59.47\% and mainly located in the bilateral precuneus. The bilateral middle temporal gyrus, and the inferior temporal gyrus, showing strong coherence with VAN, Vis, and SMN.

\begin{figure}[ht]
	\centering
	\includegraphics[trim=0 0 0 0,scale=1.0,width=12cm]{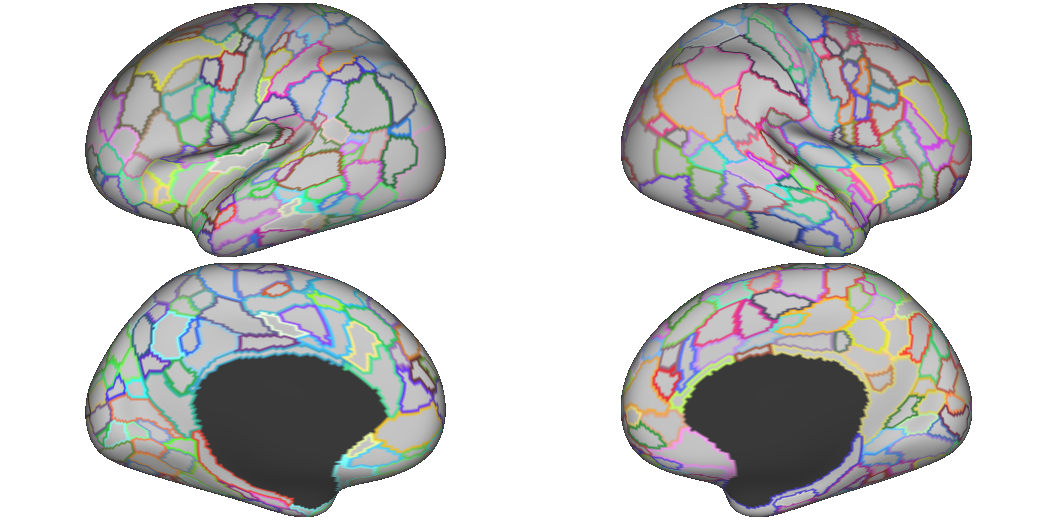}
	\caption{Final parcellation  with 456 parcels using all frequency-specific parcellations across frequencies.}
	\label{fig:merge}
\end{figure}

A final parcellation was acquired using spatial clustering of all frequency-specific parcellations with reference to \cite{Yan-2019-p-116363-116363}. As shown in Figure \ref{fig:merge}, the final parcellation has a total of 456 parcels, with 233 parcels in the left hemisphere and 223 in the right hemisphere.

\subsection{Repeatability experiment}
\begin{figure}[ht]
	\centering
	\includegraphics[trim=0 170 0 0,scale=1.0,width=12cm]{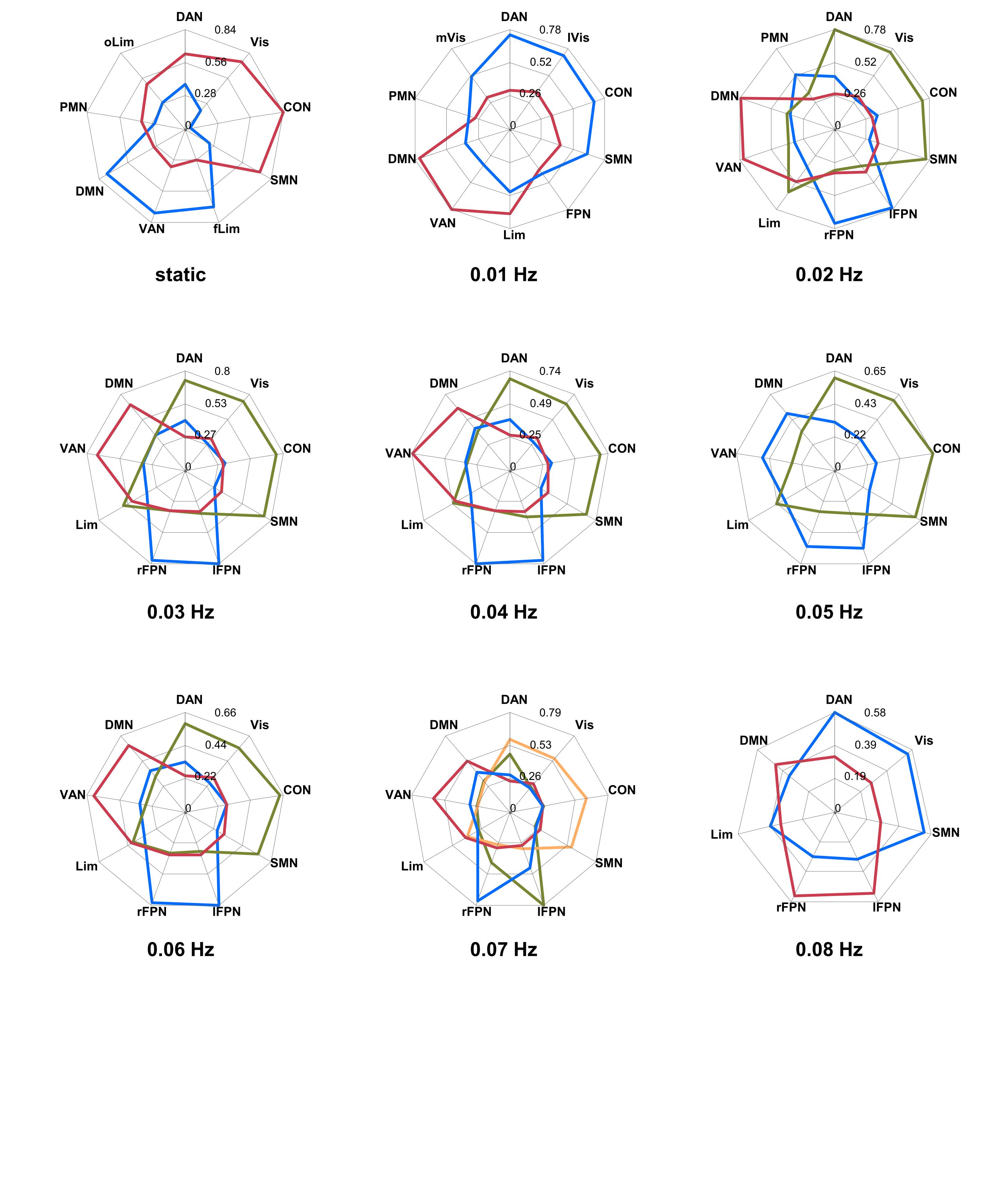}
	\caption{Repeatability experiment results using 3T fMRI data. Except for 0.05 $Hz$, the type and network connectivity mode of hubs in 3T data are completely consistent with 7T data.}
	\label{fig:3T}
\end{figure}

Repeatability experiments were conducted on the HCP 3T fMRI dataset. The functional network and parcels in the HCP 7T data were invoked as templates, and the hub detection and clustering were performed on the HCP 3T individual data. Results were shown in Figure \ref{fig:3T}. It can be seen that, the hub types and network connection modes of 3T data are completely consistent with 7T data except for 0.05 $Hz$. In the hub clustering results of the 3T data at 0.05 $Hz$, the hubs (light green line) that have strong coherence with DAN, Vis, CON, and SMN are consistent with the 7T data, while the other type of hubs (light blue line) show strong coherence with DMN, VAN, rFPN, and lFPN at the same time, which can be regarded as a combination of two types of hubs (light red and light blue line) in the 7T data at 0.05  $Hz$. 
 The repeatability experiments indicate that the type of hubs and their network connection mode at specific frequencies are the essential attributes of the human brain, rather than a random phenomenon.

\subsection{Topology of individual functional network}
\begin{figure}[ht]
	\centering
	\includegraphics[trim=0 0 0 0,scale=1.0,width=12cm]{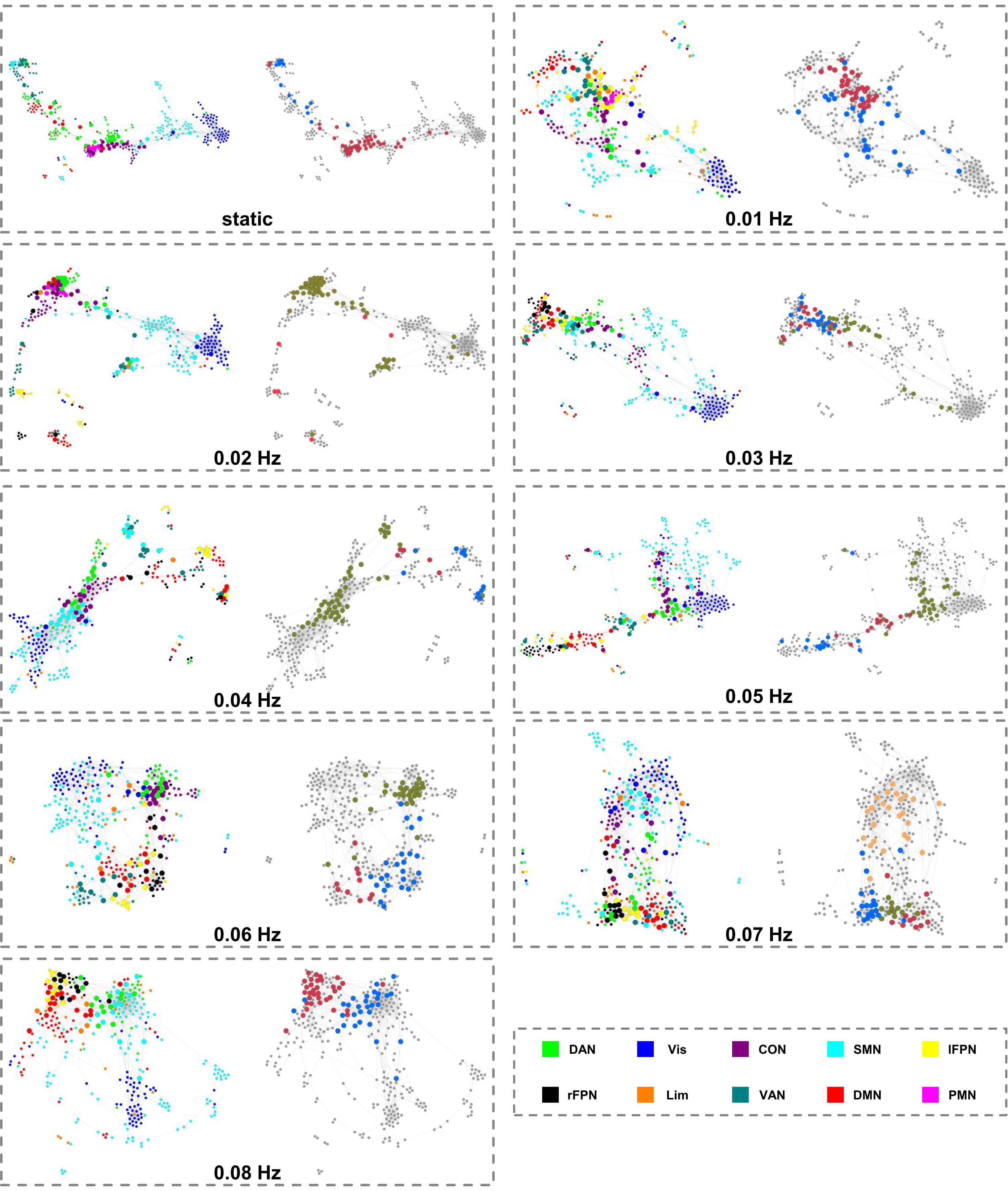}
	\caption{Spring embedding plot of individual functional network structure and hubs (ID: 100610, REST1). Except for the lower right corner, the color in the left picture in each dotted box represents the functional network, the color in the right picture represents the hub set, and the dot with a larger radius represents the hub. The point with a node degree of 0 is not displayed, and the network color label is in the lower right corner.}
	\label{fig:topology}
\end{figure}

Figure \ref{fig:topology} shows the spring embedding plot of the functional network structure of a single subject (ID: 100610) with the edge density of 2.5\%. In the spring embedding plot, the closer the relationship between nodes is, the closer their position is \citep{Power-2011-p665-678}. The spring embedding plot shows that individual functional networks do exhibit different topological structures at different frequency bands. A typical example is in figure \ref{fig:topology}, the main body of the static network has the following linear structure:
\[VAN \to DMN \to DAN \to PMN \to CON \to SMN \to Vis \]
While at 0.06 $Hz$, the main body of the functional network has the following ring structure:
\[ Vis \leftrightarrow SMN \leftrightarrow DMN \leftrightarrow VAN \leftrightarrow lFPN \leftrightarrow rFPN \leftrightarrow CON \leftrightarrow DAN \leftrightarrow Vis\]
There are also mixed linear and circular structures. For example, at 0.05 $Hz$, the main structure of the functional network is:
\[rFPN \to lFPN \to DMN \to VAN \to DAN \leftrightarrow Vis \leftrightarrow SMN \leftrightarrow CON \leftrightarrow DAN\]

\begin{figure}[ht]
	\centering
	\includegraphics[trim=0 0 0 0,scale=1.0,width=12cm]{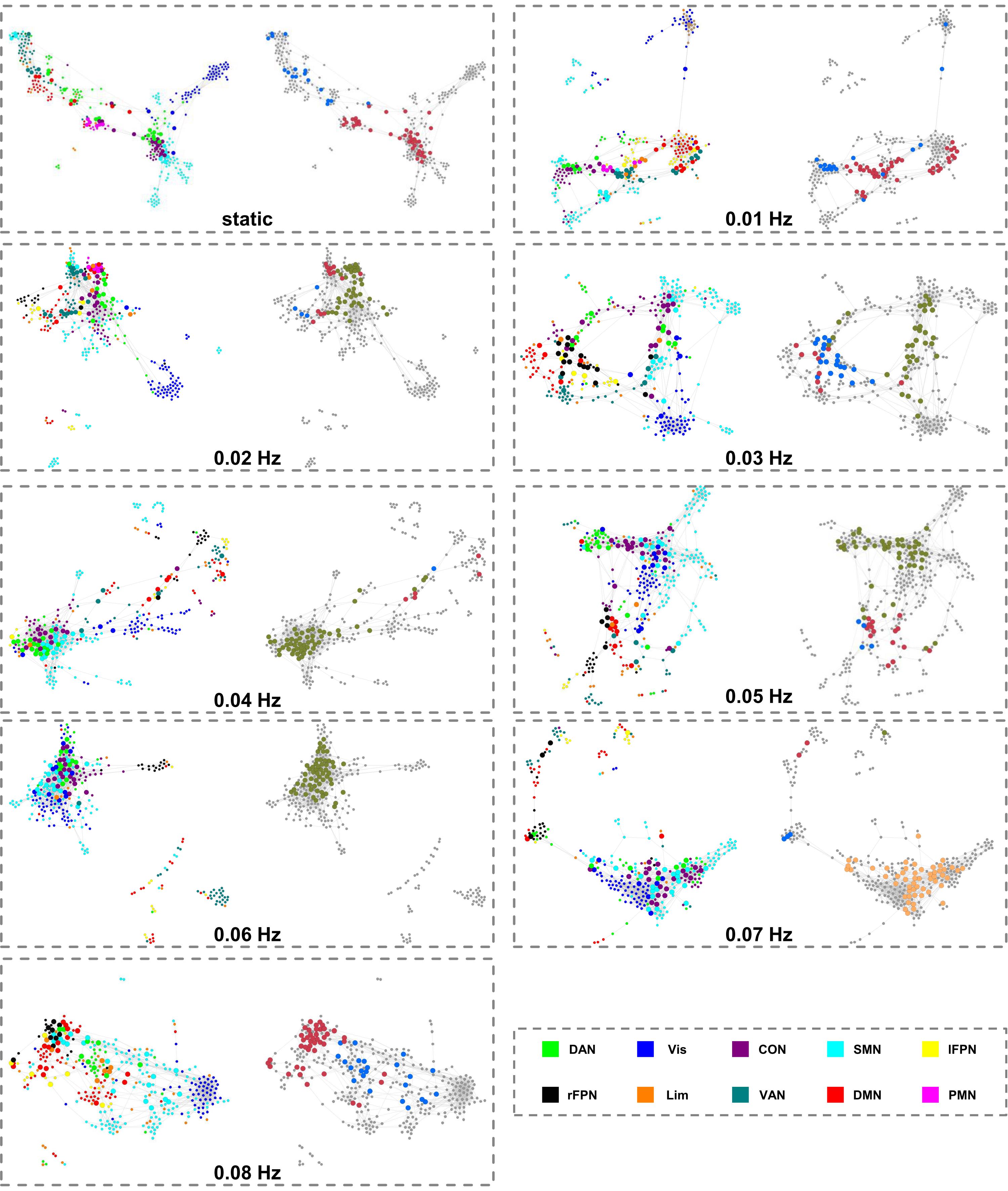}
	\caption{The intra-individual consistency between the functional network structure and the hub set (ID: 100610, REST3). Except for the lower right corner, the color in the left picture in each dotted box represents the functional network, the color in the right picture represents the hub set, and the dot with a larger radius represents the hub. The point with a node degree of 0 is not displayed, and the network color label is in the lower right corner.}
	\label{fig:topology2}
\end{figure}

We can learn from Figure \ref{fig:topology} that Vis and SMN are closely related, but they are basically located at the edge of the main body of the functional network, and hubs rarely appear in these two networks. DMN, CON, and VAN are basically located in the center of the functional network, which means that they are deeply involved in the neural activities of the resting brain. Lim is rarely included in the main body of the functional network, which may imply that Lim rarely participates in the information transmission of the resting-state functional network. The difference in the topological structure of the functional network means that the information transmission path is different across frequencies. Figure \ref{fig:topology2} shows the spring embedding plot of the functional network structure when the edge density is 2.5\% based on another scan data (REST3) of the same subject (ID: 100610). The topology structure of the static network is highly consistent across scans, while the topology structure of the frequency-specific networks is slightly different across frequencies. For example, the functional network corresponding to REST1 tends to show a linear structure at 0.03 $Hz$, while the functional network corresponding to REST3 shows an obvious ring structure, and vice versa at 0.07 $Hz$. Another example is the functional network at 0.06 $Hz$, there are 3 types of hubs corresponding to REST1, while only one type of hubs corresponding to REST3.

\section{Discussion}
In this study, we analyzed the functional parcellation of the cerebral cortex and the topology of its functional networks at different frequencies. We found that the neural activity of the human brain is frequency-specific, and we also verified the stability of the network topology at specific frequencies. Our study may provide a new perspective to the understanding of the functional human brain. Detailed discussions are as following.
\subsection{Frequency specificity of the functional parcellation}
In previous studies of resting-state functional parcellation of the cerebral cortex, static parcellation is the mainstream \citep{Luo-2019-p-269-282, Fan-2014-p3365-3378, Zhang-2014-p719-727, Shi-2018-p1358-1368, Gordon-2016-p288-303, Barnes-2012-p1148-1158, Schaefer-2018-p3095-3114, Craddock-2011-p1914-1928, Kahnt-2012-p6240-6250, Karachi-2012-p9396-9401, Blumensath-2013-p313-324, Parisot-2016-p68-83, Shen-2013-p403-415}, supplemented by dynamic parcellation \citep{Zhong-2019-p165-175, Ji-2016-p954-967}. As far as we know, there is no study on the functional parcellation of the human brain from the perspective of frequency so far. Our research shows that the functional parcellation and the way of integration and segregation of the functional networks are frequency-specific. For example, the visual network is subdivided into the lateral and central parts at 0.01 $Hz$, while this division does not exist at other frequency bands and static situations. For another example, the sensory-motor network and the cingulo-opercular network are merged into one network at 0.08 $Hz$, which is also observed in other situations. It should be noted that we cannot simply regard the merger of frequency-specific parcellation at the considered effective frequency bands (0.01$\sim$0.08 $Hz$) as further subdivisions based on static parcellation. For instance, in the static parcellation, the limbic network is subdivided into the fronto-limbic network and occipito-limbic network, while in the frequency-specific parcellations, the limbic network is not further subdivided. Therefore, the relationship between static parcellation and frequency-specific parcellation should be more complicated, which is necessary to be further studied.

\subsection{Frequency specificity of the functional network topology}
\cite{Sasai-2014-p1022-1022} explored the topological structure of the functional human brain network from the perspective of frequency. Except for this, there is a lack of research on the functional network from the perspective of frequency. We believe that the premise of correctly describing the frequency-specific topological structure of a functional network is the rationality and reliability of the frequency-specific parcellation of the cerebral cortex. 

It is mentioned that if the nodes cannot completely and accurately represent the functional units of the functional network, the established graph model must be distorted, and the topological properties of the model will not correctly reflect the real properties of the functional network \citep{ Wig-2011-p126-146, Smith-2011-p875-891, Butts-2009-p414-416}. \cite{Sasai-2014-p1022-1022} directly selected ROIs from the resting human brain network as the nodes of the frequency-specific network graph, the premise of which is that these ROIs are the smallest functional units at each frequency. In this study, we paralleled the cortex at each frequency of 0.01 $\sim$ 0.08 $Hz$ and found that the functional units and their relationships are various across frequencies.

\subsection{Clustering of connector hubs at specific frequencies}
In functional networks, a connector hub cannot have strong functional connectivities with all networks. There should be differences in the network connectivity modes of different connector hubs, which are rarely noticed before  \citep{Gordon-2018-p4-1695}. Clustering of the connector hubs helps to understand the segregation and integration of the functional human brain networks more intuitively. In this study, we observed stable frequency-specific clustering of connector hubs and depicted their modes of functional connectivities with functional networks. This is very important for us to better understand how cortical areas to work together at different frequencies.

\subsection{Limitations}
We noticed some shortcomings of the work. First, parcel creation is somewhat subjective, and the combination of functional networks and local parcel is relatively rigid. Subjectivity has always been a difficult point in functional parcellation. Although we have avoided parameter selection during network parcellation and thereby avoiding subjectivity to the greatest extent, when local parcellations were conducted, the edge density threshold which decided the boundary of parcels is chosen empirically. Although this is a general approach based on gradient subdivision  \citep{Gordon-2016-p288-303,Power-2011-p665-678,Gordon-2018-p4-1695}, it indeed affected the final parcellation results. Second, we focus on the differences of functional network topology at different frequencies in this study, so the group-average parcellation is used as the template. However, differences in individual brain functions have attracted more and more attention \citep{Gordon-2017-p918-939, Seitzman-2019-p-}, and it is necessary to consider individual differences in the future.

\section*{Funding}
	This work was supported by the National Natural Science Foundation of China (62036013, 61722313, 61773391).

\bibliography{myref}

\begin{figure}[ht]
	\centering
	\includegraphics[trim=0 50 0 0,scale=1.0,width=12cm]{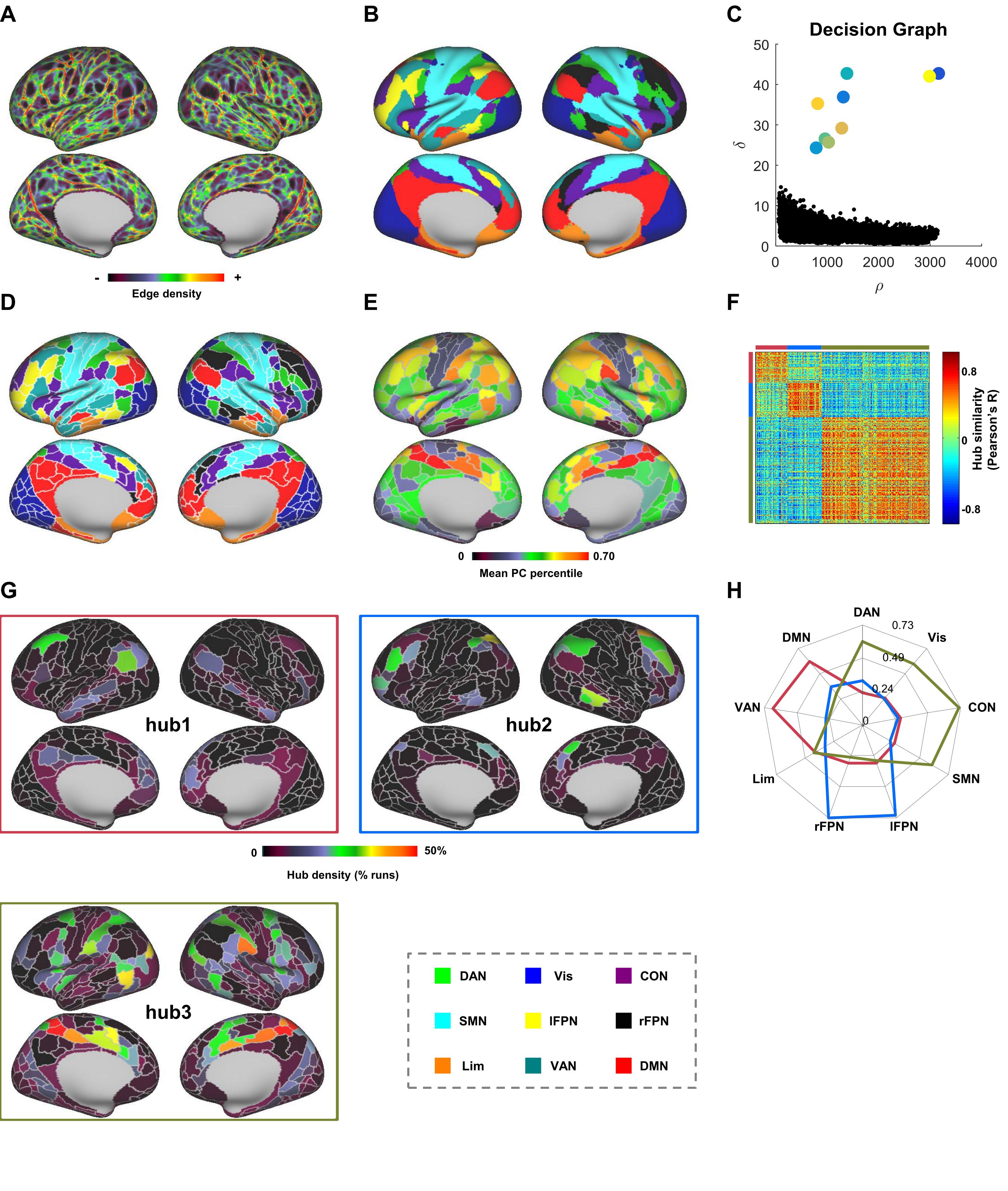}
	\caption{Cortical parcellation and hub clustering at 0.04 $Hz$. (A) Group average edge density map based on coherence at 0.04 $Hz$; (B) Eigen-clustering network parcellation results; (C) Eigen-clustering decision graph; (D) Whole cortical parcellation results; (E) Average PC percentile; (F) Hub clustering result, the border color indicates the hub set; (G) The spatial distribution of the hub set and the border color represents the hub set; (H) Average network connectivity of each set of hubs. The line color indicates the hub set.}
	\label{fig:fre4}
\end{figure}

\begin{figure}[ht]
	\centering
	\includegraphics[trim=0 50 0 0,scale=1.0,width=12cm]{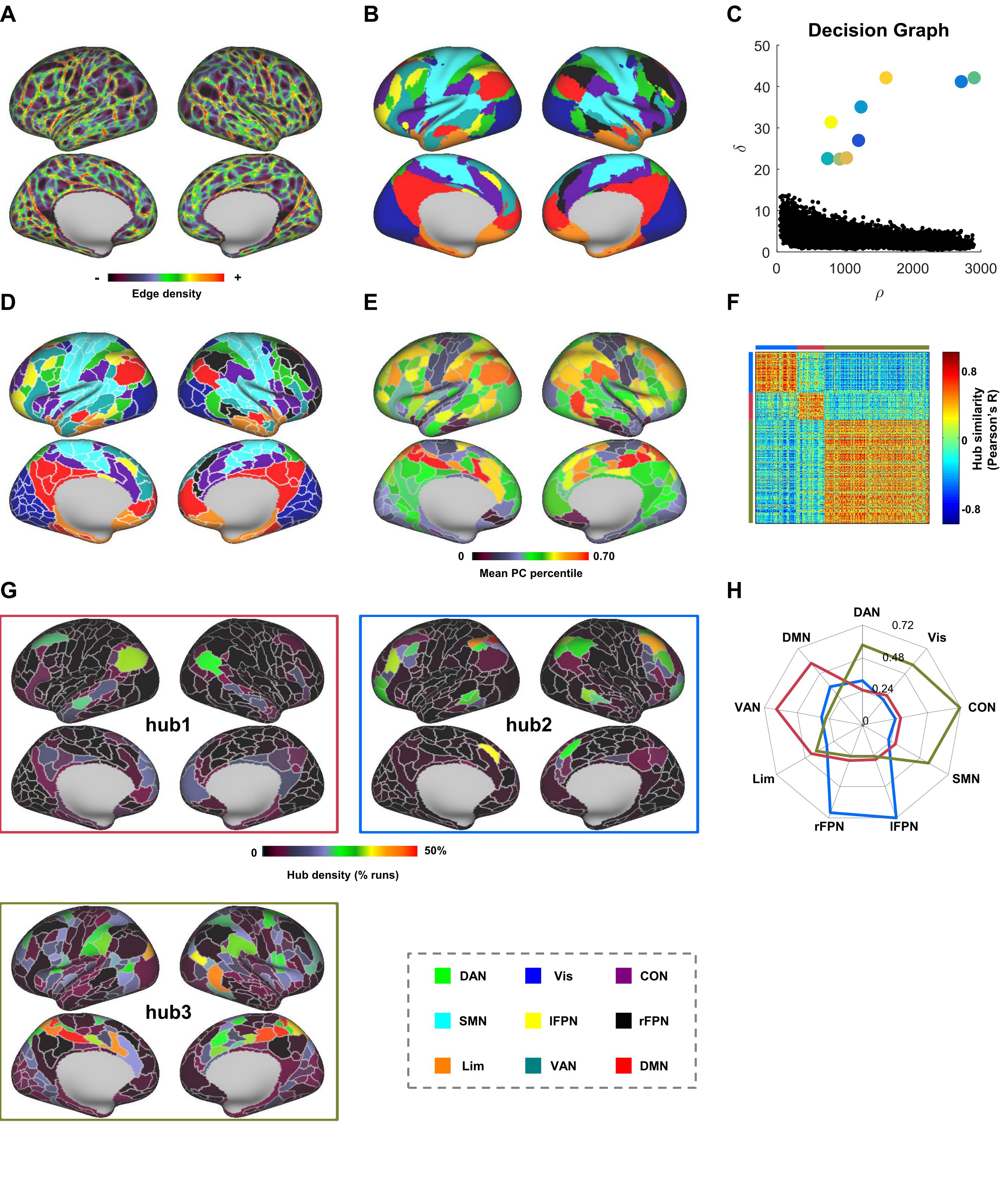}
	\caption{Cortical parcellation and hub clustering at 0.05 $Hz$. (A) Group average edge density map based on coherence at 0.05 $Hz$; (B) Eigen-clustering network parcellation results; (C) Eigen-clustering decision graph; (D) Whole cortical parcellation results; (E) Average PC percentile; (F) Hub clustering result, the border color indicates the hub set; (G) The spatial distribution of the hub set and the border color represents the hub set; (H) Average network connectivity of each set of hubs. The line color indicates the hub set.}
	\label{fig:fre5}
\end{figure}

\end{document}